\documentclass[%
aip,
amsmath,amssymb,
reprint,nofootinbib%
]{revtex4-2}
\usepackage{nomencl}
\makenomenclature
\renewcommand\nomgroup[1]{%
  \item[\bfseries
  \ifstrequal{#1}{A}{Nomenclature}{%
  \ifstrequal{#1}{N}{Abbreviations}{%
  \ifstrequal{#1}{B}{Subscripts}{%
  \ifstrequal{#1}{C}{Superscripts}{%
  \ifstrequal{#1}{D}{Greek Letters}{%
  \ifstrequal{#1}{X}{Other symbols}{}}}}}}%
]}

\usepackage{graphicx,array,booktabs,rotating,parskip}
\usepackage{dcolumn}
\usepackage{bm}
\usepackage{mathptmx}
\usepackage{comment}
\usepackage{enumitem}
\usepackage{afterpage,float,color,xcolor}
\usepackage{multirow}
\usepackage{lipsum} 
\usepackage{hyperref}
\usepackage{etoolbox} 
\usepackage{tabularx,tabulary}
\usepackage[capitalize]{cleveref}
\usepackage{times}

\hypersetup{
	colorlinks,
	linkcolor={blue!100!black},
	citecolor={blue!100!black},
	urlcolor={blue!100!black}
}

\makeatletter
\newcommand\footnoteref[1]{\protected@xdef\@thefnmark{\ref{#1}}\@footnotemark}
\makeatother
\renewcommand{\figurename}{FIGURE}

\usepackage[page]{totalcount}
\usepackage{etoolbox,fancyhdr,xcolor}%
\usepackage{ragged2e}
\usepackage[displaymath,pagewise]{lineno}

\begin{document}
\preprint{AIP/123-QED}
\title[ Manuscript under preparation $|$ 2026]{Hypersonic Flow Over an Axisymmetric Open Cavity — Effects of Segmentation Angle and Reynolds Number}

\author{Talluri Vamsi Krishna}
\email{sri.vamsi1432@gmail.com}
\affiliation{Technion Wind Tunnels Complex, Faculty of Aerospace Engineering, Technion-Israel Institute of Technology, Haifa-3200003, Israel}
\author{Soumya R. Nanda}
\email{srnanda@iitj.ac.in  (corresponding author)}
\affiliation{Department of Aerospace Engineering, Indian Institute of Technology Jodhpur, Karwar-342030, Rajasthan, India}

\author{S. K. Karthick}
\email{skkarthick@mae.iith.ac.in}
\affiliation{Department of Mechanical and Aerospace Engineering, Indian Institute of Technology Hyderabad, Kandi-502285, Telangana, India}

\author{Jacob Cohen}
\email{aerycyc@gmail.com}
\affiliation{Technion Wind Tunnels Complex, Faculty of Aerospace Engineering, Technion-Israel Institute of Technology, Haifa-3200003, Israel}


\begin{abstract}
This study investigates the influence of cavity segmentation angle  ($\phi$) on the flow characteristics of a cavity mounted on a cone under Mach 6 freestream conditions in a Ludwieg tunnel. Four segmentation configurations, corresponding to $\phi = 360^\circ$, $48^\circ$, $32^\circ$, and $16^\circ$, are examined for cavity 
length-to-depth ratios ($L/D$) of 1 and 6 over a range of Reynolds numbers ($Re_D = 23000$--$74000$). The flow field is characterized using surface pressure and heat flux measurements, complemented by planar laser Rayleigh scattering visualization. The results indicate that, for $L/D = 1$, the downstream Strouhal 
number ($St$) and mean pressure remain largely unaffected by the segmentation angle and closely agree with those of the axisymmetric configuration. In contrast, the flow parameters within the cavity are larger than in the axisymmetric case, which is attributed to the three-dimensional bounding effect. For $L/D = 6$, the mean pressure distribution exhibits a similar trend across the segmentation angles, whereas the downstream $St$ increases with increasing $\phi$, approaching the axisymmetric value. Flow-mode transitions are observed across all segmentation angles, consistent with the axisymmetric configuration. However, both the transition Reynolds numbers and the corresponding flow modes vary with the segmentation angle.

\end{abstract}
 
\maketitle


\nomenclature[A]{$L$}{Cavity length (m)}
\nomenclature[A]{$D$}{Cavity depth (m)}
\nomenclature[A]{$p$}{Instantaneous static pressure (Pa)}
\nomenclature[A]{$T$}{Instantaneous static temperature (K)}
\nomenclature[A]{$M$}{Mach number }
\nomenclature[A]{$u$}{Velocity (m/s)}
\nomenclature[A]{$k$}{Thermal conductivity (W/m·K)}
\nomenclature[A]{$Re$}{Reynolds number per m ($\rho u /\mu$, 1/m)}
\nomenclature[A]{$t$}{Time (ms)}
\nomenclature[A]{$f$}{Frequency (Hz)}
\nomenclature[A]{$n$}{Mode number}
\nomenclature[A]{$\rho$}{Density (kg/m$^3$)}
\nomenclature[A]{$\mu$}{Dynamic viscosity (kg/ms)}
\nomenclature[A]{$\theta$}{Semi-apex angle ($^\circ$)}
\nomenclature[A]{$S$}{Streamwise distance along the shear layer (m)}
\nomenclature[A]{$\theta$}{Semi-apex angle ($^\circ$)}
\nomenclature[A]{$\phi$}{Segmentation angles ($^\circ$)}
\nomenclature[A]{$c$}{Specific heat capacity}
\nomenclature[A]{$\beta$}{Effective thermal product}

\nomenclature[B]{$\infty$}{Freestream quantity}
\nomenclature[B]{$0$}{Total or stagnation quantity}
\nomenclature[B]{$2$}{Quantities incoming to the cavity}
\nomenclature[B]{$D$}{Based on cavity depth}
\nomenclature[B]{$r$}{Rear face}
\nomenclature[B]{$fc$}{Floor center}
\nomenclature[B]{$T$}{CSJT}
\nomenclature[B]{$p$}{Pressure Sensor}
\printnomenclature


\section{Introduction} \label{sec:intro}
Cavities can present on aerospace vehicles, either intentionally designed or unintentionally formed, across a wide range of flow regimes, from subsonic to hypersonic speeds. They are used mainly in applications such as landing gear, internal weapon bays, scramjet and ramjet flame holding, flow control, and sensor housing, and they also occur unintentionally at thermal panel gaps, control surface regions, and fastener and bolt recesses. Generally, the flow past cavities is complex due to interactions among the shear layer, shock waves, and recirculation within the cavity. Previous studies suggest that the flow over the cavity depends on the Mach number and the geometrical properties of the cavity length (L) and depth (D)\cite {rossiter1962effects}. Based on the shear-layer impingement location, Rossiter \cite{rossiter1962effects} classified the cavity into three categories: open ($L/D\leq10$), closed ($L/D\geq14$), and transitional ($10\geq L/D\leq13$) based on the $L/D$ ratio.

 Krishnamurthy\cite{krishnamurty1955acoustic} initially studied the flow over rectangular cavities from subsonic to low supersonic speeds and observed the Mach-number dependence of cavity resonance tones and their acoustic radiation. Rossiter\cite{rossiter1964wind} observed the self-sustained oscillations in the shear layer over the cavity. Later, Heller and Bliss\cite{heller1975andBliss} analyzed the shear layer over open cavities to explain the cause of flow unsteadiness. They proposed that under steady conditions, the shear layer should contain a stagnation point at the rear wall. Due to shear effects, the stagnation streamline changes direction, causing the flow to strike the rear wall at an angle. For a cavity with a vertical rear wall, this would require the shear layer to curve significantly over the cavity starting. Such curvature leads to static pressure variations across the cavity mouth that cannot be balanced by the internal cavity pressure, resulting in shear-layer instability. Conversely, if the shear layer remains straight to avoid pressure imbalance, the required impingement angle cannot be maintained, making the flow near the rear stagnation point inherently unsteady. The shear layer impinging at the rear wall of the cavity produces acoustic waves that propagate upstream within the cavity. They also observed in supersonic cavity flows that periodic mass supply and removal from the cavity are responsible for acoustic radiation. When these waves reach the leading edge, they disturb the flow as it separates from the front wall, triggering additional oscillations in the shear layer. This process establishes a continuous feedback loop \cite{rossiter1964wind}.  Tam et al. \cite{tam1996} performed computational simulations and observed that a vortex impinging at the cavity rear wall generates a pressure pulse that drives an upstream-traveling wave at the shedding frequency. This wave triggers new vortex formation at the leading edge. Although the initial reflected wave dissipates, a secondary wave forms beneath the vortex, reflects from the aft corner, and sustains the feedback cycle. Rossiter\cite{rossiter1964wind} has proposed a semi-empirical formula to calculate the cavity-associated resonance frequency, later modified by Heller et al.\cite{Heller1971}, which is presented in terms of the Strouhal number $(St_D)$ in the equation 
 \ref{eq:rossiter},    
\begin{align}
\label{eq:rossiter}
\left [\frac{fD}{u}\right] =\frac{(n-\alpha)(D/L)}{M_\infty\left(1+\left[(\gamma-1)/2\right]M_\infty^2\right)^{-1/2}+1/k},
\end{align}
Here, M denotes the Mach number of the flow upstream of the cavity, while  n=[1,2,3,4,…] represents the integer mode number. In this study we have used, the empirical constants  $\alpha =0.25$ and $k=0.57$ correspond to the phase lag in the feedback loop and the ratio of the convection velocity of cavity shear layer to the freestream velocity, respectively \cite{Murray2001, rossiter1964wind}.Unalmis et al.\cite{Unalmis2001} studied cavity flows at Mach 5; they observed that, at hypersonic Mach numbers, acoustical phenomena inside the cavity have negligible effect on shear-layer dynamics and suggested that the lack of coupling between the shear layer and acoustics leads to purely acoustic modes. Nicoll\cite{NICOLL1964} studied axisymmetric laminar cavity flows at Mach 11 and observed at hypersonic speeds laminar cavity flow is very stable similar to the attached boundary layer. In case of $L/D \leq7$ the floor pressure is constant for 90\% of cavity length and of $L/D \geq7$ pressure gradient exists on the cavity floor and recompression takes place along the entire cavity length. Heat transfer on the cavity floor is reduced to 10\% to 20\% of the attached-flow conditions. 

%

However, depending on the cavity's geometry, the separated flow often transitions to turbulence, increasing stagnation heat transfer and imposing high pressure on nearby components. Therefore, understanding the unsteady characteristics of the separated shear layer under various flow conditions is crucial during the design phase. In addition, the intended or unintended cavity placement may be axisymmetric (combustor) or occupy a certain segment of the base body (sensor housing, TPS tiles, etc.), rather than a fully axisymmetric configuration.  In line with the perspective mentioned earlier, the present experiments are planned to study the effect of $L/D$ ratios in axisymmetric as well as varying the segmentation ($\phi$) of the cavity for $[L/D] = [1,6]$, will be assessed interms of the change in  flow field behaviour and heat transfer rate for a wide range of Reynolds numbers ($Re_D$). 
\section{Experimental Methodology} \label{sec:exp_meth}

\begin{figure*}
	\includegraphics[width=0.8\textwidth]{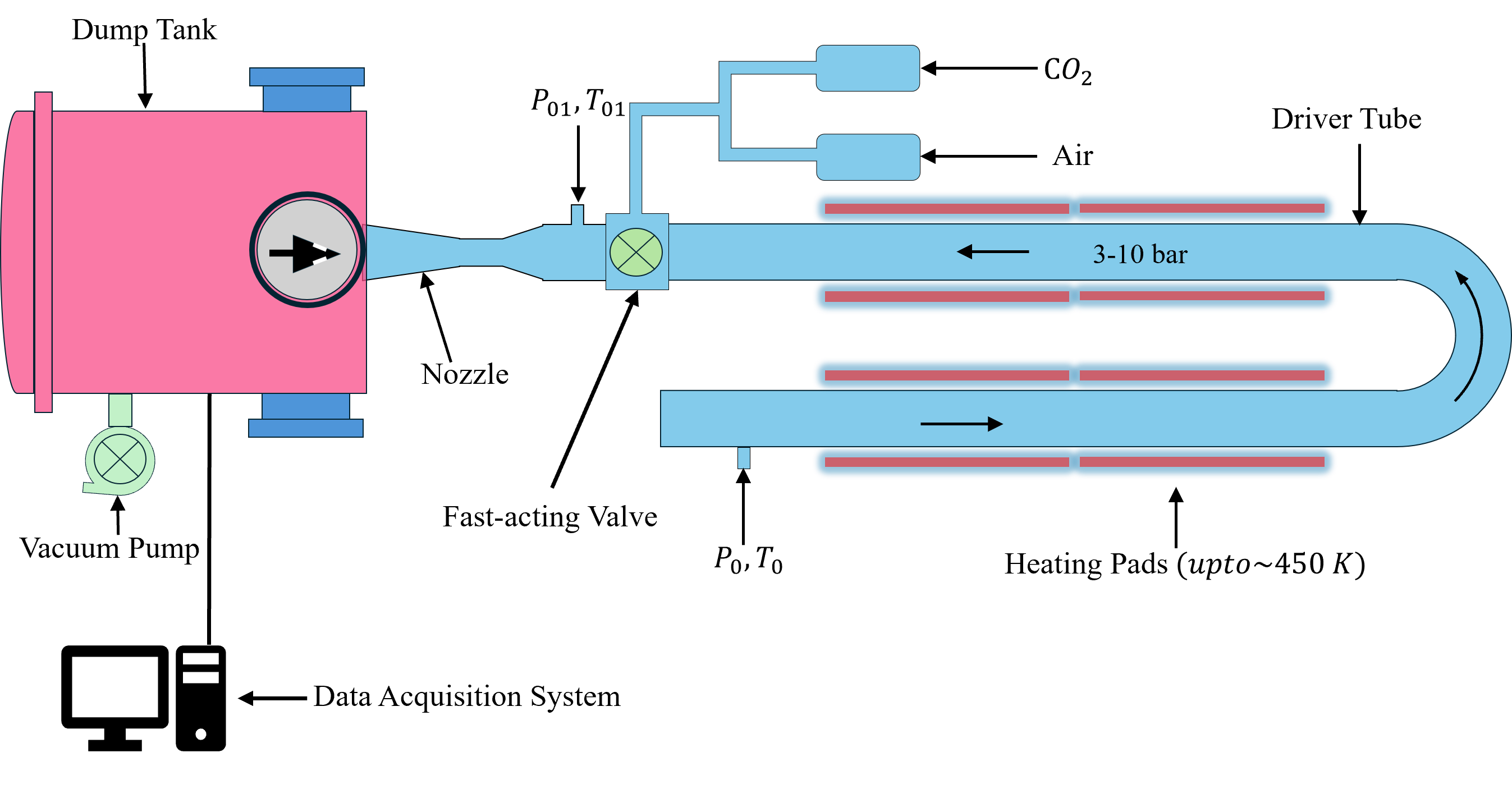}
	\caption{Schematic of present experimental facility}
    \label{fig:Facility}
\end{figure*}

\begin{figure*}
    \includegraphics[width=0.8\textwidth]{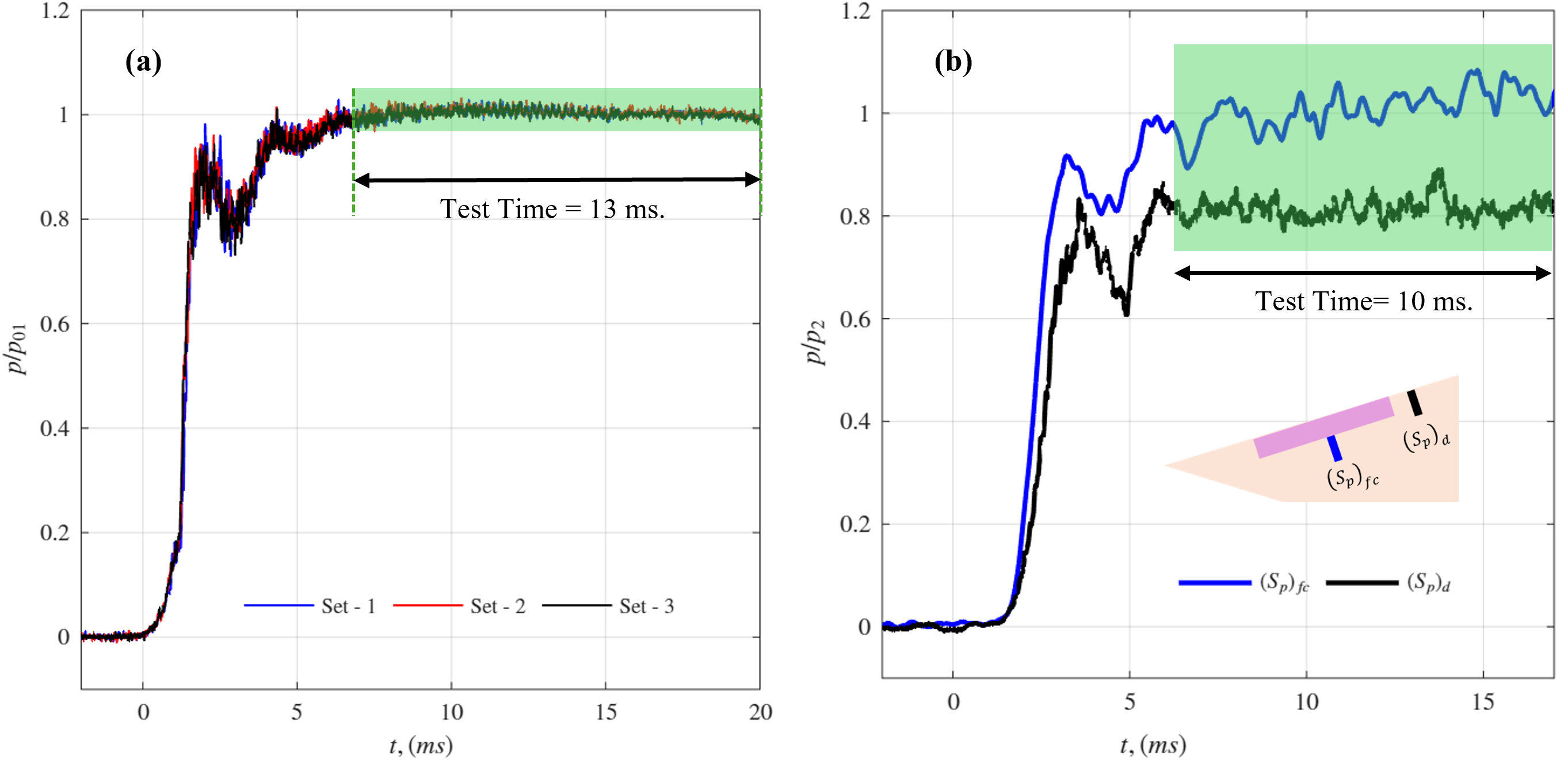}
	\caption{\label{fig:press_sample} (a) Pressure signal repeatability measured upstream of the C–D nozzle. (b) Sample signals obtained from the pressure sensors mounted on the cavity floor: $(S_p)_{fc}$, and downstream of the cavity: $(S_p)_{d}$ (See Figure \ref{fig:model_details}c for sensor location details) for the case of $[L/D,  Re_D] = [6.0, 74000]$.  }
\end{figure*}

Present experiments are carried out in the Hypersonic Ludwieg Tunnel (HLT)(see Figure \ref{fig:Facility}) at Technion Wind Tunnels Complex (TWTC), Technion-IIT, Israel. The HLT comprises a $3$ m long driver tube in a `U'-tube configuration. The tube is filled with compressed air at a desired fill pressure between $3$ and $10$ bar to produce a wide range of Reynolds numbers based on cavity depth ($Re_D$). The driver tube is separated from the convergent-divergent (C-D) nozzle and test section through a fast-acting valve (ISTA\textsuperscript{\textregistered} KB-20-10) having a response time of $1$ ms. Typical pressure losses across the fast-acting valve are around 20\%.  A vacuum pressure of $0.05$ mbar is maintained in the test section to generate the pressure ratio needed to draw a hypersonic free jet flow of $M_\infty=6.0$ across the nozzle exit. The steady test window is determined from the stagnation pressure $(p_{01})$ signal recorded upstream of the C–D nozzle. The $p_{01}$ time history from \ref{fig:press_sample}-a delineates the interval during which the reservoir conditions remain nearly constant. During operation, a steady flow is achieved for $13$ ms. The core flow of the nozzle is estimated at about 60\% of the nozzle exit diameter. Readers are directed to a previous publication \cite{Karthick_2023} to learn more about the facility, operational methods, and flow characterization. Table \ref{tab:freestream_cond} presents extreme cases of freestream conditions generated in the test section.

\begin{table*}
\caption{The estimated flow parameters in the freestream and cavity upstream for the extreme cases of storage fill pressure.}
\label{tab:freestream_cond}
\begin{ruledtabular}
\begin{tabular}{ccc}
\textbf{Parameters} & \textbf{Freestream (${\infty}$)} & \textbf{Cavity upstream ($2)$}\\
\hline
Static pressure ($p$, Pa) & 152.64 - 514.92 & 1472 - 4785\\
Static temperature ($T$, K) & 36.58 & 91.6\\
Static Density ($\rho$, kg/m$^3$) & 0.015 - 0.049 & 0.056 - 0.182\\
Kinematic viscosity ($\nu \times 10^{-5}$, m$^2$/s) & 14.7 - 4.4 & 11.2 - 3.4\\
Velocity ($u$, m/s) & 727.46 & 636.93\\
Mach number ($M$) & 6.0 & 3.32\\
Reynolds number ($Re_D \times 10^3$, $D$: cavity depth) & 19 - 67 & 23 - 74\\
\end{tabular}
\end{ruledtabular}
\end{table*}

\subsection{Model details} \label{ssec:model_det}
\begin{figure*}
	\includegraphics[width=1\textwidth]{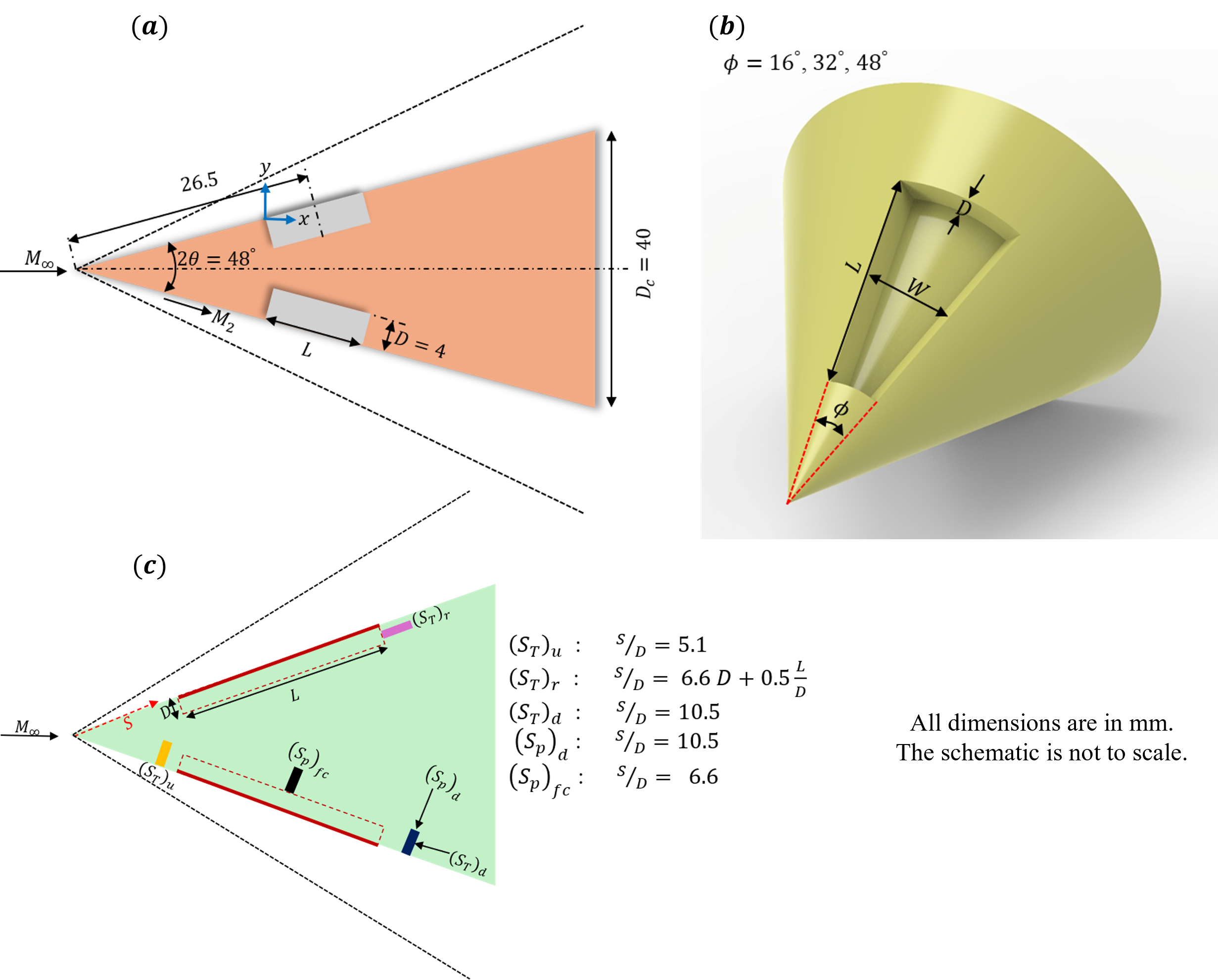}
	\caption{\label{fig:model_details} Geometrical variations of the (a) axisymmetric (b) segmented cone-cavity model (c) pressure sensors and CSJTs locations used for the axisymmetric and segmented cavity configurations.}
\end{figure*}

A cone having a base diameter of $D_c=40$ mm and an apex angle of $2\theta=48^\circ$ is considered. Different axisymmetric cavity models are made in rectangular grooves with a constant depth of $D=4$ mm. The overall length of the cone is kept constant at $45$ mm. The aspect ratio (length-to-depth ratio, $L/D$) of the cavity is varied as $[L/D]=[1.0,6.0]$. In all the cases, the cavity center is kept at a constant distance of $26.5$ mm from the cone tip (See Figure \ref{fig:model_details}-a).  Furthermore, $3$-D cavities having segmentation angles of $\phi=16^\circ, 32^\circ, 48^\circ$  (equivalent $W/D = 0.5, 1, 2$) is studied for the both L/D configurations (See Figure \ref{fig:model_details}-b). Experiments are carried out on the cone-cavity configuration as shown in Figure \ref{fig:model_details} for a set of upstream cavity flow Reynolds numbers of $[Re_D]=[23,000;\ 30,000;\ 38,000;\ 45,200;\ 51,300;\ 59,000;\ 66,300;\\ 74,000]$ (based on the cavity depth). The freestream and upstream flow conditions are detailed in Table \ref{tab:freestream_cond}. All the experiments are performed three times to ensure repeatability.

\subsection{Model instrumentation details} \label{ssec:instrument}

This study primarily aims to quantify the mean pressure and heat loading in different zones of the cone-cavity models. Unsteady pressure sensors (Kulite\textsuperscript{\textregistered} XCQ-080-100A) are mounted on the cavity floor center ($(S_p)_{fc}$), and at a location downstream the cavity, $(S_p)_d$ ($S/D = 10.5$) as shown in Figure \ref{fig:model_details}-c .To estimate the local heat flux, coaxial surface junction thermocouples \cite{heufer2008} (CSJT) (RWTH-Aachen-HGD\textsuperscript{\textregistered} KLP and MI-3.6) are installed at the rear face center ($(S_T)_r$), and downstream of the cavity ($(S_T)_d$), for all the cases in consideration. Moreover, due to space limitations, only for $L/D = 1$, a CSJT is mounted upstream of the cavity ($(S_T)_u$) at $S/D = 5.1$ to evaluate the local heat flux from the attached incoming flow to the cavity (See Figure \ref{fig:model_details}-c). All the sensor signals are recorded using the NI data acquisition cards at 0.8 MHz for the pressure sensors and 100 kHz for the thermocouples. Sample pressure signals obtained from the sensors mounted on the cavity floor, viz. $(S_p)_{fc}$ and  cavity downstream ($(S_p)_{d}$) for the case of $[L/D] = [6.0]$ at $Re_D = 74000$ is shown in Figure \ref{fig:press_sample}-b.  The steady-state time of these sensors during the $10$ ms runtime is observed and used for further analysis.

\subsection{Visualization techniques} \label{ssec:visual_tech}

As part of this study, planar laser Rayleigh scattering (PLRS) flow visualization imaging is performed for both the axisymmetric and segmented cone–cavity configurations to obtain the required qualitative and quantitative flow information. Measurements in a particular streamwise plane (at $z/D=0$) are carried out using the PLRS technique \cite{Do_2010, Zhang_2016}, and detailed information on the present PLRS setup is provided in \cite {Karthick_2023}. For the segmented cavity configuration, the shear layer could not be adequately resolved when the camera was positioned perpendicular to the laser sheet. Therefore, to maintain uniformity across all test cases and to ensure proper visualization of the shear layer, the camera was tilted by 10°. This modification was implemented by leveraging the characteristic isotropic scattering behavior of light in the Rayleigh scattering regime, which allows reliable imaging even when the camera is not positioned strictly perpendicular to the laser sheet\cite{Miles2001LaserRS}. A 10 W continuous 532 nm laser with a 5 mm beam diameter from a CNI\textsuperscript{\textregistered} laser (MGL-V-532-10W) is used as the illumination source. The scattered light intensity is captured using a Phantom high-speed camera(T4040). High-frequency PLRS images are acquired at a resolution of $256\times 128$, with an acquisition rate of 0.2 MHz. 

\begin{figure*}
	\includegraphics[width=0.8\textwidth]{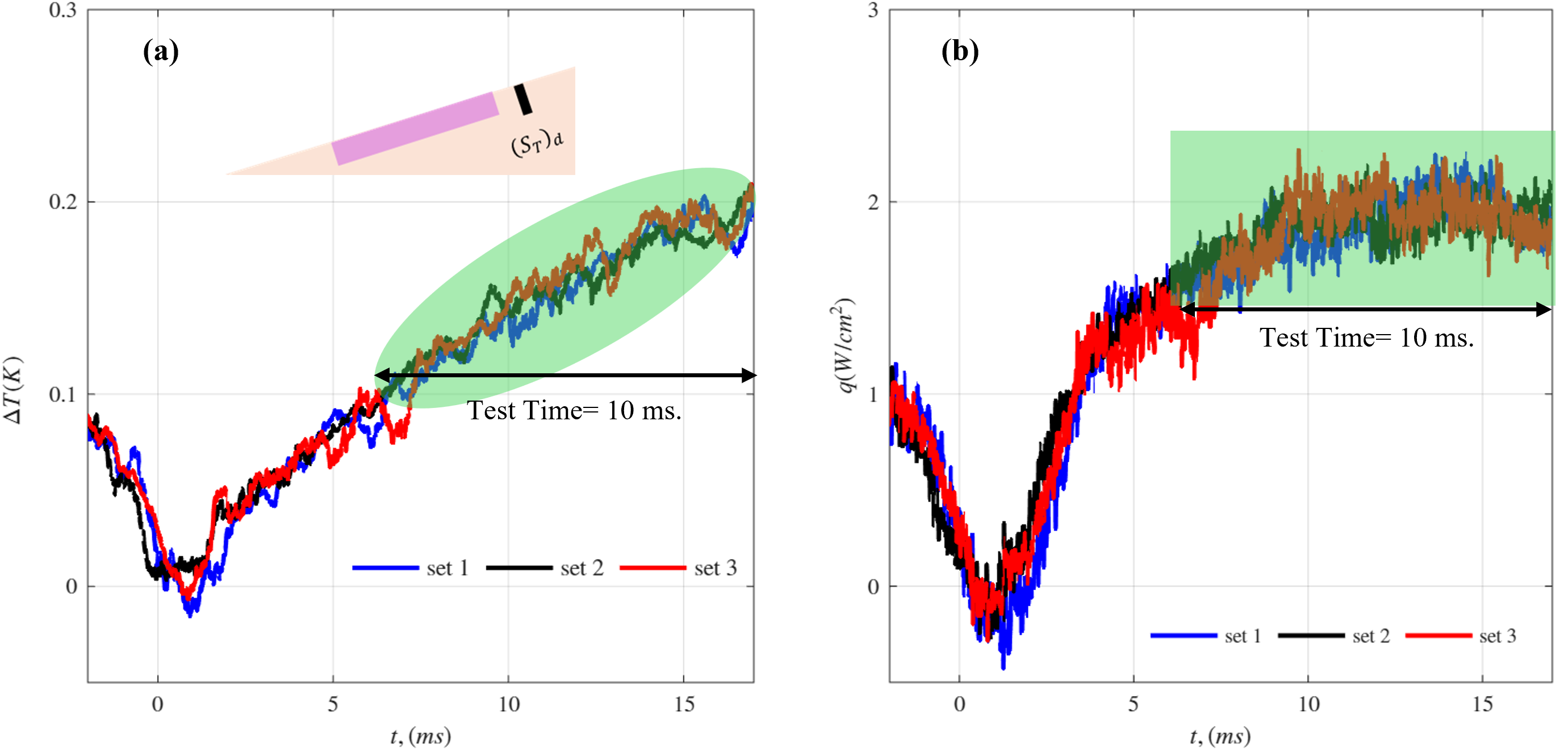}
	\caption{Sample (a) rise in temperature and (b) heat flux signal for three runs obtained from the CSJT mounted at the cavity downstream location: $(S_T)_d$, for the case of $[L/D, Re_D] = [6.0,  74000]$. The green colour horizontal marking demarcates the steady heat flux value during which HF is considered to be constant.}
    \label{fig:Repeatability HF}
\end{figure*}

\subsection{Heat Flux Estimation} \label{ssec:HF_est}

The local heat flux is estimated from the temperature rise ($\Delta T$) measured using an E-type CSJT. The temperature rise at the downstream cavity location ($(S_T)_d$) is shown in Figure \ref{fig:Repeatability HF}a along with the repeatability of the responses. The surface heat flux ($\dot{q}_s$) is calculated using the one-dimensional heat conduction equation following the analytical procedure proposed by Hollis\cite{hollis1995user}, as given by the following equation \ref{eq:ob_eqn}.

\begin{align}
\label{eq:ob_eqn}
\dot{q_s} =& \frac{2\beta}{\sqrt{\pi}} \sum_{i=1}^{n} \frac{T(t_i) - T(t_{i-1})}{{{(t_n-t_i)}^{1/2}} + {{(t_n-t_{i-1})}^{1/2}}} ;\, \beta =\sqrt{\rho c k},
\end{align}

where $\rho$ is the density, $c$ is the specific heat capacity, $k$ is the thermal conductivity of the substrate material, and $\beta$(8500 $J/m^2KS^{0.5}$) is defined as the "effective thermal product (TP)" of the substrate material.

In response to the parabolic rise in temperature (Figure \ref{fig:Repeatability HF}-a), the local heat flux was estimated using Equation \ref{eq:ob_eqn} and is shown in Figure \ref{fig:Repeatability HF}-b. The calculated heat flux values are consistent across all three runs, supporting confidence in the measured temperature rise during the test. The steady heat flux values, marked in green in Figure \ref{fig:Repeatability HF}-b, are used for the comparative assessment. To generalize the wall heat flux values, the Stanton number is estimated using Equation \ref{eq:stanton} shown below.

\begin{align}
\label{eq:stanton}
St = \left[\frac{\dot q_s}{\rho_2 u_2 (h_{2}-h_w)}\right] ; \ h_{2}=C_p\times T_{2},   \ h_{w}=C_p\times T_{w}
\end{align}
Where the subscripts $_2$ and $_w$ represent cavity upstream and wall properties, respectively. 
\section{Results and Discussion}

This section emphasizes the quantitative and qualitative findings from the sensors (pressure and CSJT) and PLRS flow visualization for the axisymmetric and segmented cavity($\phi=16^\circ, 32^\circ, 48^\circ$) configurations. The flow features observed in PLRS imaging across all cavity configurations are shown.  The time-varying light-intensity fluctuations at a point ($x/D= 0.9 \frac{L}{D}$) on the shear layer are subjected to a Fast Fourier Transform (FFT) to identify the dominant frequencies, which are then compared with the sensor-based measurements.

\subsection{Effect of segmentation angle}

\subsubsection{PLRS Visualization}
\begin{figure*}
	\includegraphics[width=0.5\textwidth]{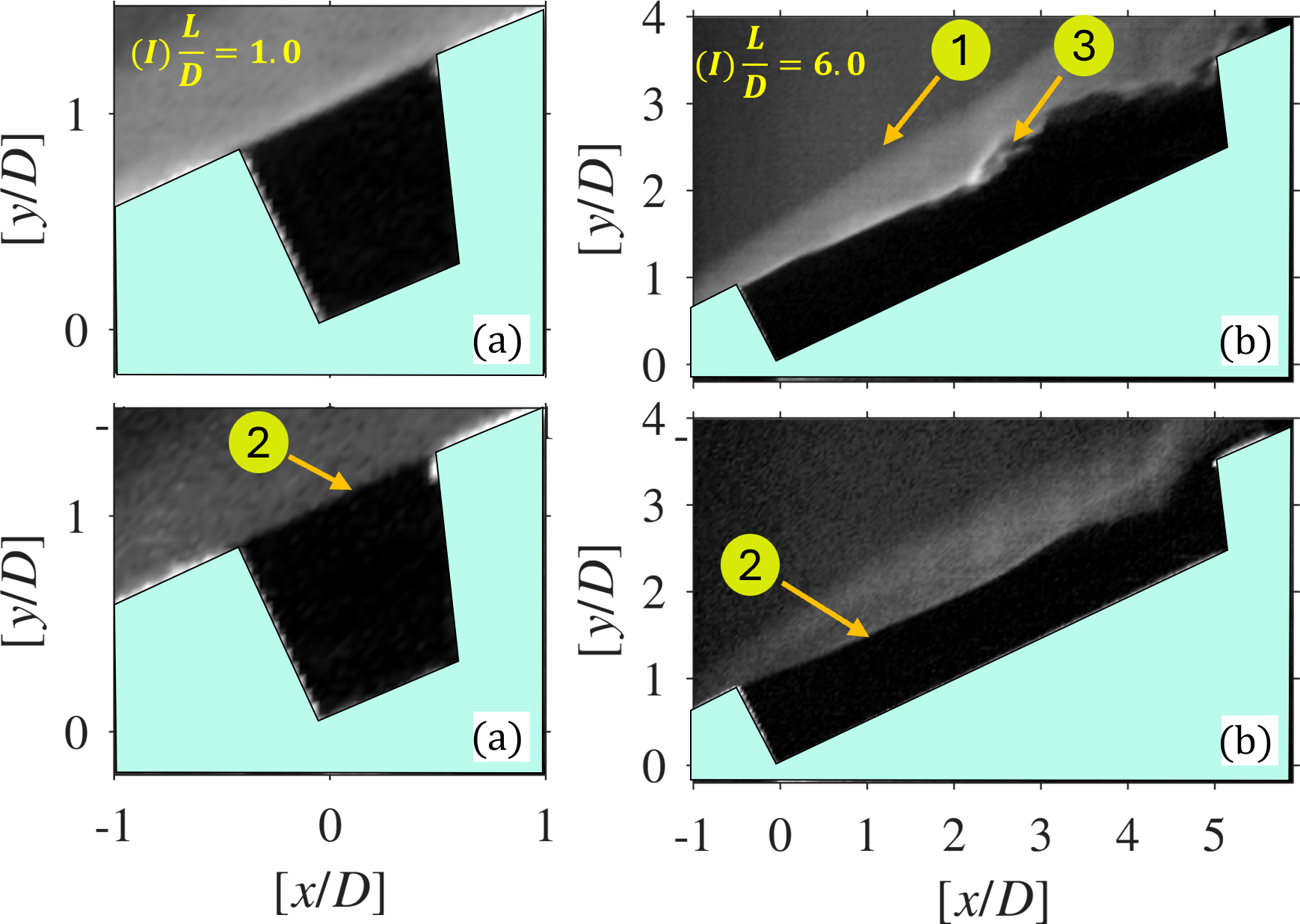}
	\caption{Instantaneous Planar Rayleigh scattering images taken at an arbitrary time step for $\phi = 360^\circ$: I. $L/D=1.0$ II. $L/D=6.0$. Different Reynolds numbers: (a) $Re_D=74000$, (b) $Re_D=38000$. Flow is from right to left. Key flow features: 1. oblique shock from the cone tip, 2. laminar separated shear layer, and 3. Kelvin-Helmholtz structures in the shear layer.}
    \label{fig:PLRS_segmentation_LD1and6}
\end{figure*}

\begin{figure*}
	\includegraphics[width=1.0\textwidth]{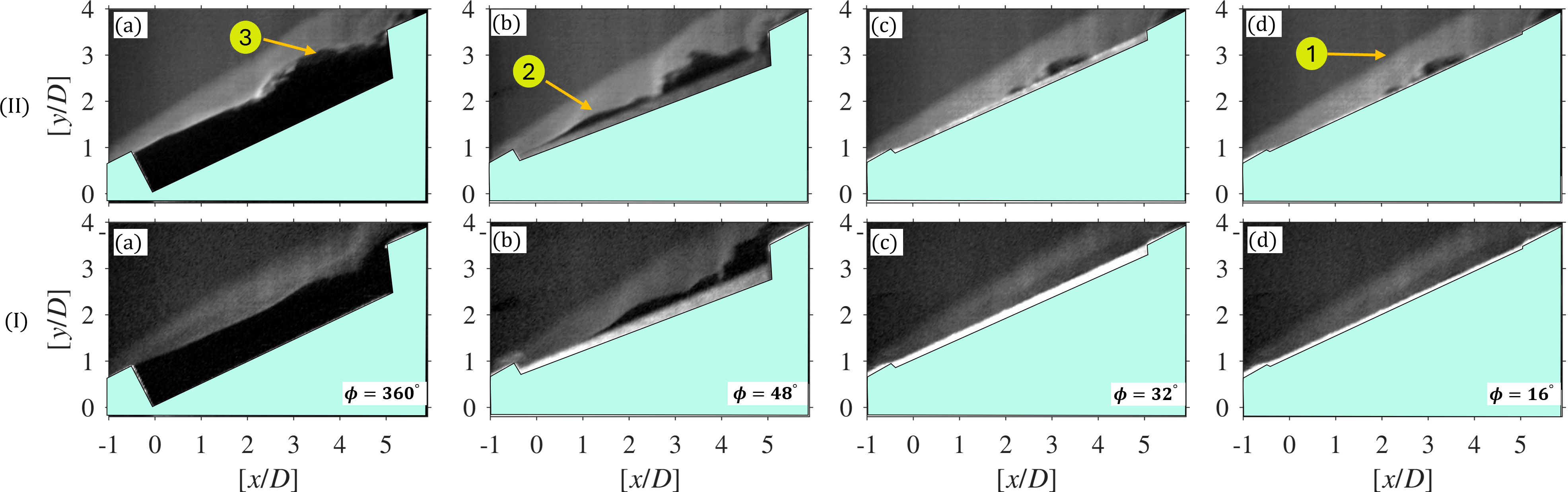}
	\caption{Instantaneous Planar Rayleigh scattering images taken at an arbitrary time step at $L/D=6$  I. $Re_D=38000$  II. $Re_D=74000$. Different segmentation angles include: (a) $\phi = 360^\circ$, (b) $\phi = 48^\circ$, (c) $\phi = 32^\circ$, and (d) $\phi = 16^\circ$. Flow is from right to left. Key flow features: 1. oblique shock from the cone tip, 2. laminar separated shear layer, and 3. Kelvin-Helmholtz structures in the shear layer.}
    \label{fig:PLRS_segmentation_LD6}
    \end{figure*}

Figure \ref{fig:PLRS_segmentation_LD1and6} shows instantaneous images obtained using Planar Laser Rayleigh Scattering $(PLRS)$ for $L/D=1.0$ and $6.0$ at $Re_D=74000$ and $38000$, respectively, for a segmentation angle $(\phi)$ of $360^\circ$. The instantaneous images reveal key flow features, including the oblique shock near the cone tip, the separated shear layer originating from the cavity leading edge, and Kelvin–Helmholtz (K–H) structures developing within the shear layer. Figure \ref{fig:PLRS_segmentation_LD1and6}-I shows the PLRS images obtained for $L/D=1.0$. The images indicate that the shear layer remains laminar at both high and low Reynolds numbers ($ Re_D$). As the cavity aspect ratio increases to $L/D=6.0$ (see Figure \ref{fig:PLRS_segmentation_LD1and6}-II), the PLRS images at the higher Reynolds number $(Re_D=74000)$ show clear K–H vortices. This indicates the transition from a laminar to a turbulent shear layer. At the lower Reynolds number ($Re_D=38000$), most of the shear layer remains laminar and impinges on the cavity rear wall.

Instantaneous PLRS images obtained for the $L/D=6.0$, at $Re_D=38000$ and 74000 for segmented angles of $\phi= 360^\circ, 48^\circ, 32^\circ$ and $16^\circ$ are  shown in Figure \ref{fig:PLRS_segmentation_LD6}. PLRS visualizations at lower $Re_D=38000$ (see Figure \ref{fig:PLRS_segmentation_LD6}-I), for smaller segmentation angles ($\phi = 16^\circ, 32^\circ$) with the camera being placed at an angle to capture the highest possible view of the shear layer, the flow within the cavity is not at all visible (see Figure \ref{fig:PLRS_segmentation_LD6}-I(c),(d)); as a result, the flow structures might not be properly captured. Nevertheless, the oblique shock originating from the cone tip is properly seen. For $\phi = 48^\circ$, the laminar part of the shear layer is clearly visible (Figure \ref{fig:PLRS_segmentation_LD6}-I(b)), whereas the axisymmetric cavity configuration depicts the shedding of K-H instability structures (see Figure \ref{fig:PLRS_segmentation_LD6}-I(a)). The instantaneous PLRS snapshots for different segmentation angle ($\phi$) at $[L/D, Re_D]=[6, 74000]$ is presented in Figure \ref{fig:PLRS_segmentation_LD6}-II. At a higher Reynolds number, the cavity with $\phi=360^\circ$ clearly showed Kelvin-Helmholtz structures in the shear layer. Similar behavior is observed for the other segmentation angles ($\phi=48^\circ, 36^\circ, 16^\circ$); increasing $Re_D$ caused the shear layer to transition.

\subsubsection{Mean pressure variation}

\begin{figure*}
	\includegraphics[width=0.8\textwidth]{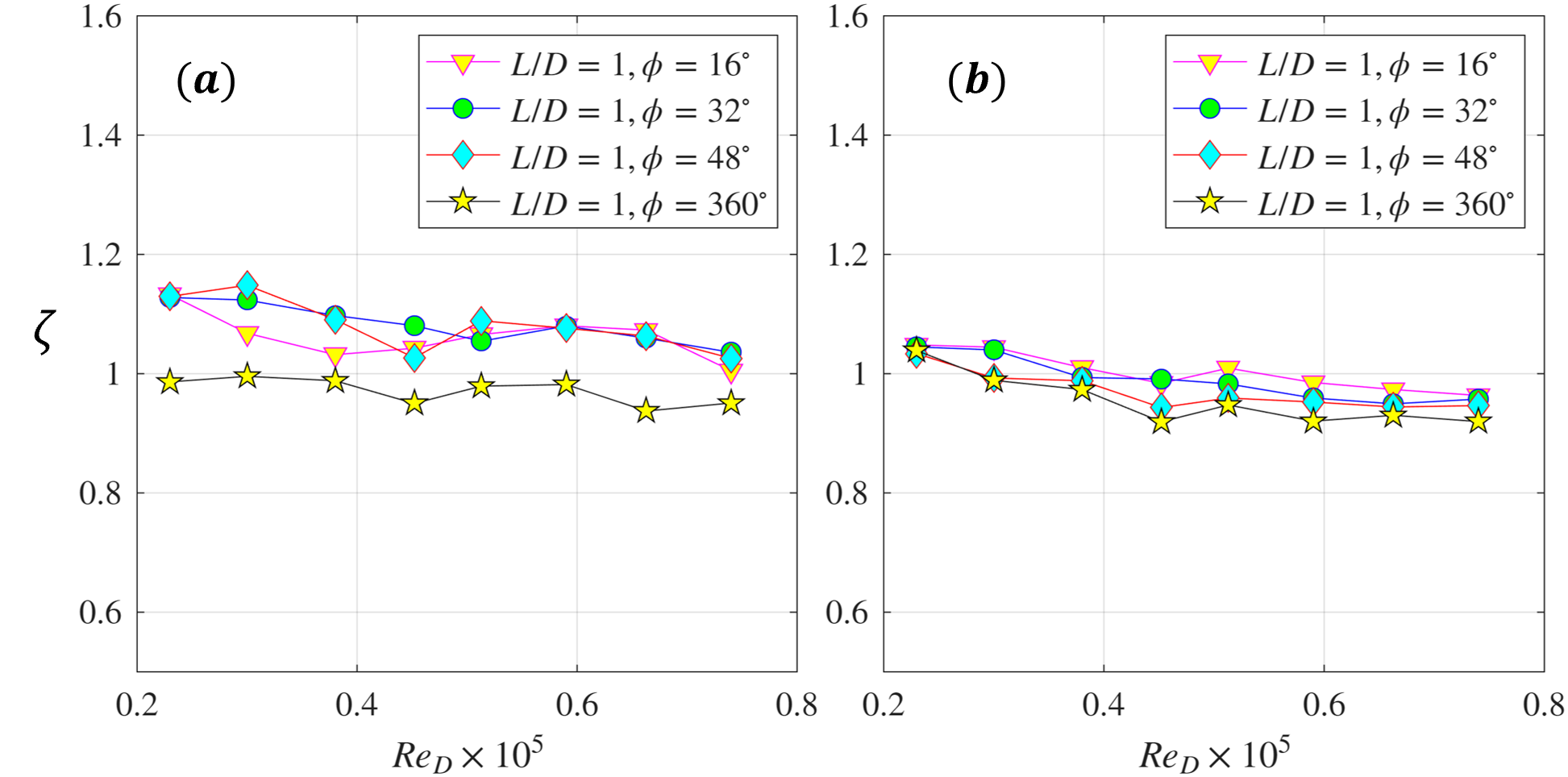}
	\caption{ The normalized mean pressure on the (a) cavity floor:$(S_p)_{fc}$ and (b) downstream cavity:$(S_p)_{d}$ location with respect to the variation in $Re_D$ and $\phi$ at $L/D = 1$.}
    \label{fig:p_phi_effect_LD1}
\end{figure*}

The variation of the non-dimensional mean pressure ($\zeta=p/p_2$) on the cavity floor ($(S_p)_{fc}$) and the downstream cavity location ($(S_p)_d$), with variation in Reynolds number ($Re_D$) and different segmentation angles ($\phi$) at $L/D=1.0$ is presented in Figure \ref{fig:p_phi_effect_LD1}. As shown in Figure \ref{fig:p_phi_effect_LD1}-a, $\zeta$ at $(S_p)_{fc}$ is consistently higher for all segmented configurations than for the corresponding axisymmetric configuration ($\phi=360^\circ$) across the entire range of $Re_D$ considered. The higher mean pressure on the cavity floor may be associated with the three-dimensional confinement or 3-D bounding effect introduced by the cavity segmentation. The presence of the bounding surfaces modifies the near-wall flow topology and the associated pressure field within the cavity, thereby potentially enhancing the mean pressure level on the cavity floor. In contrast, the mean pressure at the $(S_p)_d$ shows comparatively weak sensitivity to the segmentation angle, as evident from Figure \ref{fig:p_phi_effect_LD1}-b. The values remain within a relatively narrow range of $0.9 \leq \zeta \leq 1.05$ for both segmented and axisymmetric configurations across the entire range of $Re_D$ investigated. This limited variation suggests that, although cavity segmentation substantially influences the pressure distribution within the cavity, its effect on the mean pressure at the downstream location is comparatively small for $L/D=1.0$.

\begin{figure*}
	\includegraphics[width=0.8\textwidth]{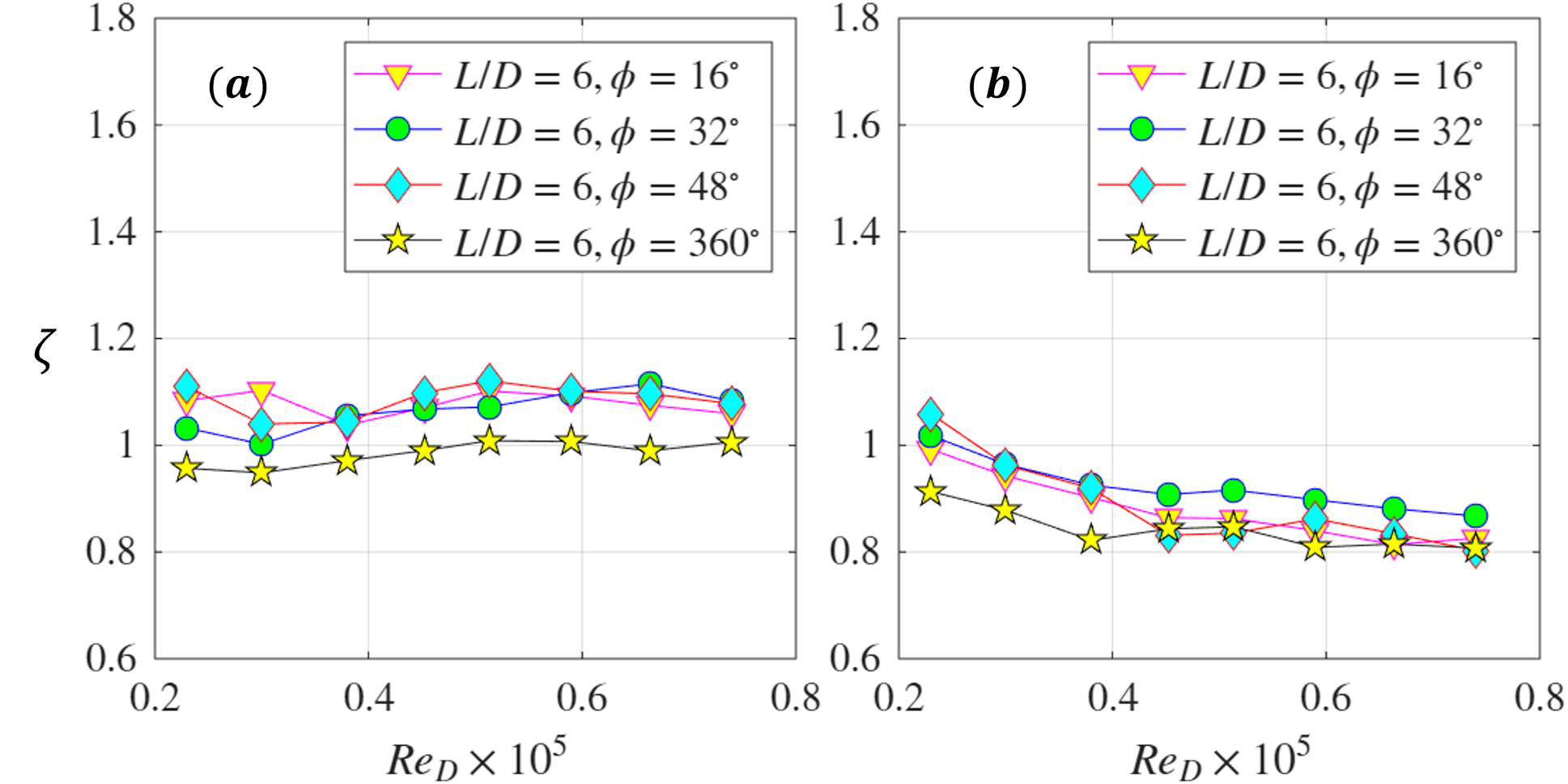}
	\caption{ The normalized mean pressure on the (a) cavity floor:$(S_p)_{fc}$ and (b) downstream cavity:$(S_p)_{d}$ with respect to the variation in $Re_D$ for different segmentation angle at $L/D = 6$.}
    \label{fig:p_phi_effect_LD6}
\end{figure*}

The normalized mean pressure ($\zeta$) obtained at the $(S_p)_{fc}$ and $(S_p)_d$ over the entire range of $Re_D$ considered for different segmentation angles ($\phi$) at $L/D=6.0$ is presented in Figure \ref{fig:p_phi_effect_LD6}. As observed in Figure \ref{fig:p_phi_effect_LD6}-a, the mean pressure at $(S_p)_{fc}$ varies similarly to the $L/D=1.0$ configuration. For all the considered segmentation angles, the segmented cavity configurations exhibit approximately $15\%$ higher mean pressure levels than the corresponding axisymmetric configuration ($\phi=360^\circ$). This consistent increase indicates that the influence of cavity segmentation on the pressure field within the cavity persists as the cavity aspect ratio increases. As $Re_D$ increases, $\zeta$ reaches 1. This is consistent with the earlier PLRS visualization (see Figure \ref{fig:PLRS_segmentation_LD6}-II): as $Re_D$ increases, the PLRS visualization clearly shows K-H vortices, indicating a laminar-to-transitional transition in the shear layer.  This likely reflects improved mixing and interaction with the freestream flow due to turbulence breakdown in the shear layer, as seen in the high-resolution schlieren imaging \cite{Nanda2026}.  

At the  $(S_p)_d$, a similar trend to that observed for $L/D=1.0$ is evident, although a modest increment in the normalized mean pressure is observed for the $L/D=6.0$ configuration, as shown in Figure \ref{fig:p_phi_effect_LD6}-b. The magnitude of this increment decreases with increasing $Re_D$, indicating that the segmentation angle has a progressively weaker influence on the downstream mean pressure at higher Reynolds numbers. Thus, while cavity segmentation produces a pronounced increase in the mean pressure at the cavity floor, its influence on the downstream pressure becomes comparatively less significant as $Re_D$ increases.

\subsubsection{Heat flux variation}

\begin{figure*}
	\includegraphics[width=0.7\textwidth]{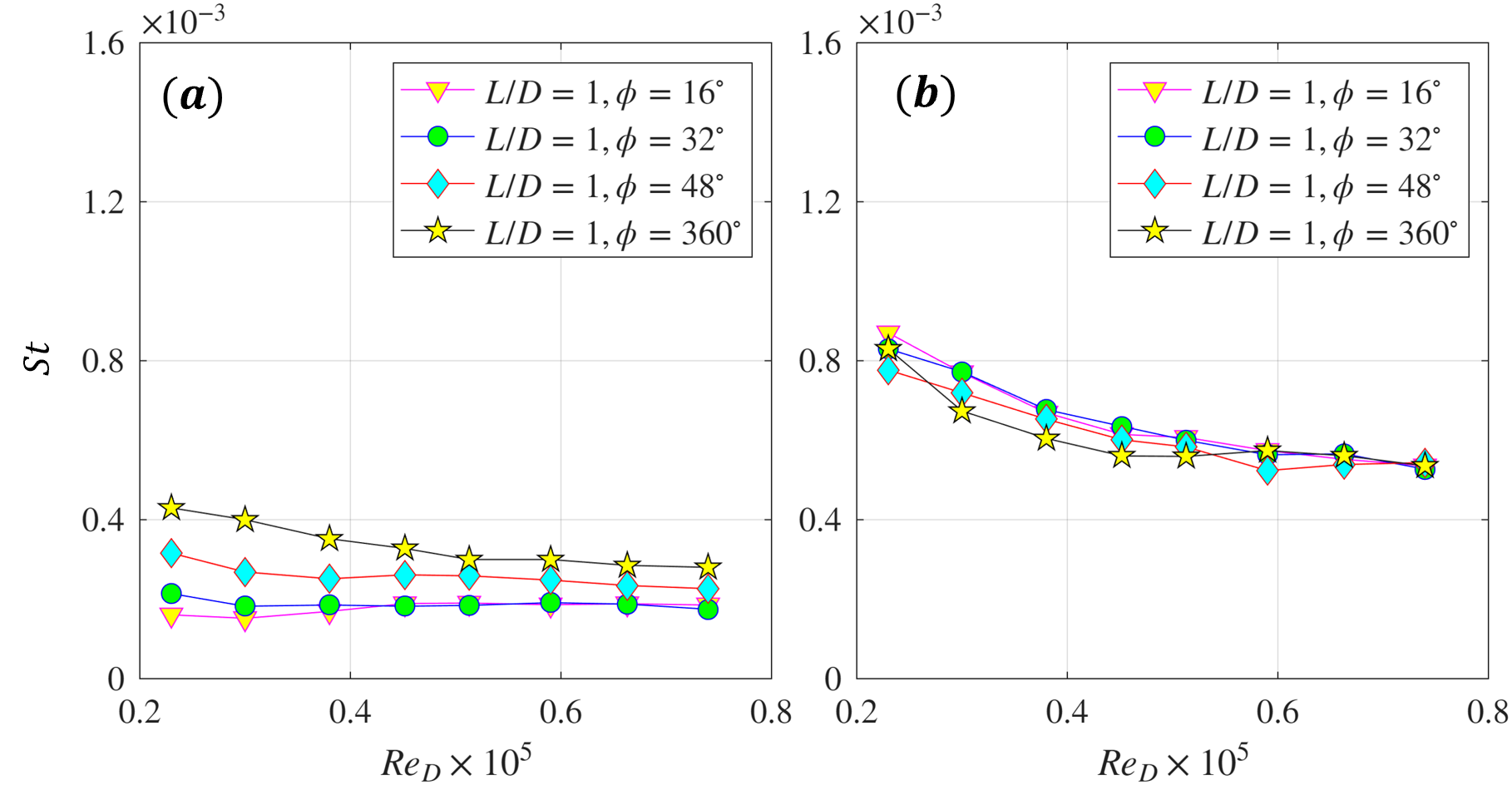}
	\caption{Variation of $St$ for (a)  cavity rear face:$(S_T)_r$,  and (b) cavity downstream:$(S_T)_d$,  location with respect to the variation in $Re_D$ and $\phi$ at $L/D = 1$.}
    \label{fig:HF_phi_effect_LD1}
\end{figure*}

The Stanton number variation for cavity aspect ratio of $L/D=1.0$, with respect to changes in $Re_D$, and segmentation angle $(\phi)$ at cavity rear wall $((S_T)_r)$ and downstream $((S_T)_d)$ locations is presented in Figure \ref{fig:HF_phi_effect_LD1}. The $St$ variation on the cavity rear wall location $((S_T)_r)$ is effected by the segmentation angle, $\phi$ (see Figure \ref{fig:HF_phi_effect_LD1}-a). The highest $St$ is observed for the axisymmetric cavity configuration, and the cavities with $\phi = 16^\circ$ and $32^\circ$ are the lowest. As $\phi$ increases, the $St$ increases for a particular $Re_D$. Also, as $Re_D$ increases, the percentage change in $St$ decreases. Similar to the mean pressure observation at the post-cavity location (see Figure \ref{fig:p_phi_effect_LD1}-b), the $St$ also does not vary significantly with the change in segmentation angle (see Figure \ref{fig:HF_phi_effect_LD1}-b). For higher $Re_D$s, $St$ remains nearly constant regardless of changes in $\phi$.

\begin{figure*}
	\includegraphics[width=0.7\textwidth]{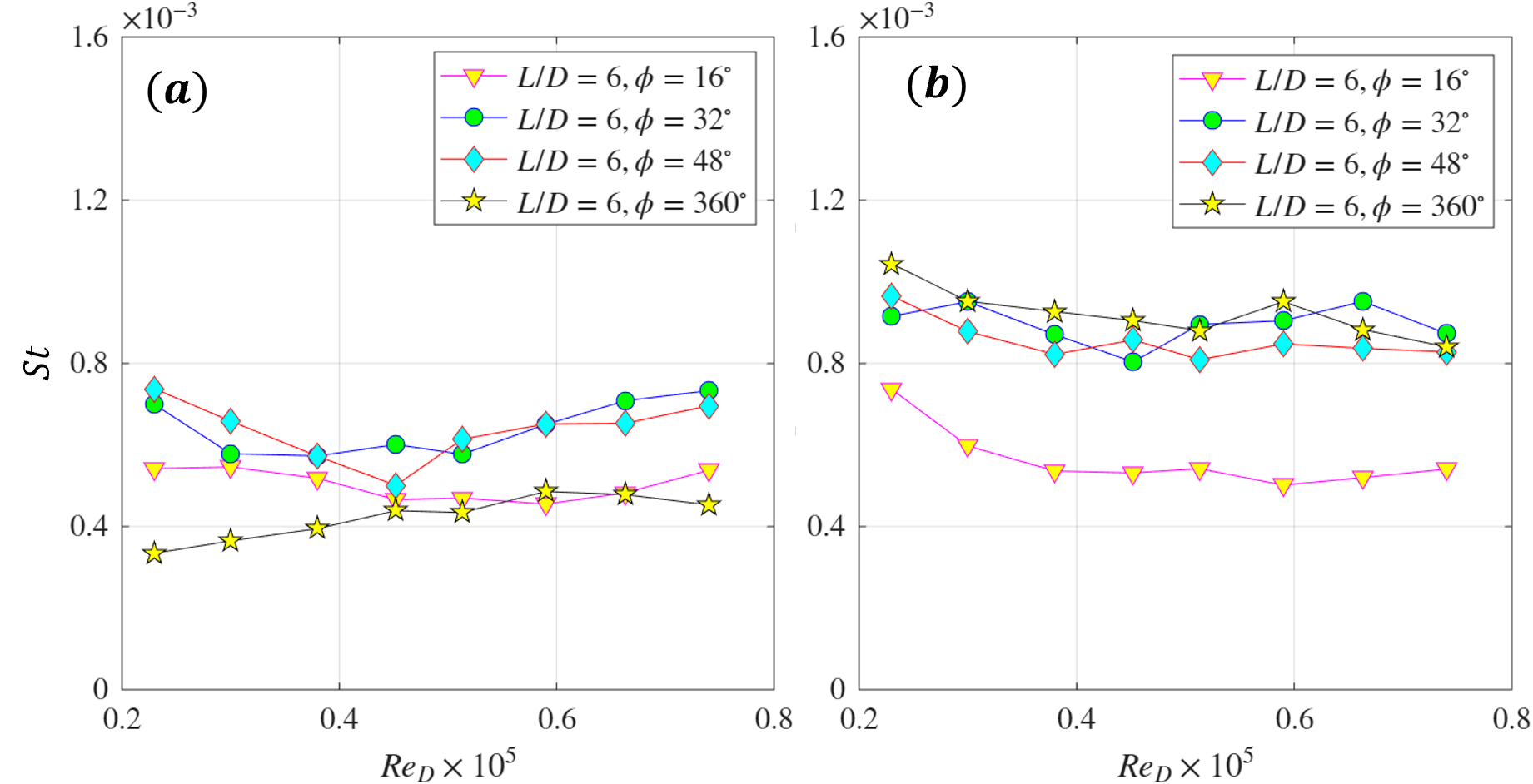}
	\caption{Variation of $St$ at (a) cavity rear face, $(S_T)_{r}$  and (b) cavity downstream, $(S_T)_{d}$, location with $Re_D$ for different $\phi$ at $L/D = 6$.}
    \label{fig:HF_phi_effect_LD6}
\end{figure*}

The $St$ variation for the $(S_T)_{r}$ and $(S_T)_{d}$ location for different segmentation angle over a wide range of $Re_D$ is shown in Figure \ref{fig:HF_phi_effect_LD6}. As shown in Figure \ref{fig:HF_phi_effect_LD6}-a, for $\phi = 16^\circ$, $St$ is much lower at the $(S_T)_{d}$ location than in other cases. As $\phi$ increases further, $St$ appears to approach the axisymmetric cavity configuration. For $\phi = 16^\circ$, the $(S_T)_{d}$ and $(S_T)_{r}$ location seems to have almost the same $St$ magnitude. Nevertheless, for other cases in consideration, the $St$ corresponding to $(S_T)_{d}$ location seems to be higher than the $(S_T)_{r}$. For the $(S_T)_{r}$ location, irrespective of the $Re_D$, the $St$ seems to be higher for the segmented cavity configurations (see Figure \ref{fig:HF_phi_effect_LD6}-b). However, no trend in $St$ as a function of $\phi$ is observed. 

\subsubsection{OASPL Variation}
\begin{figure*}
	\includegraphics[width=0.65\textwidth]{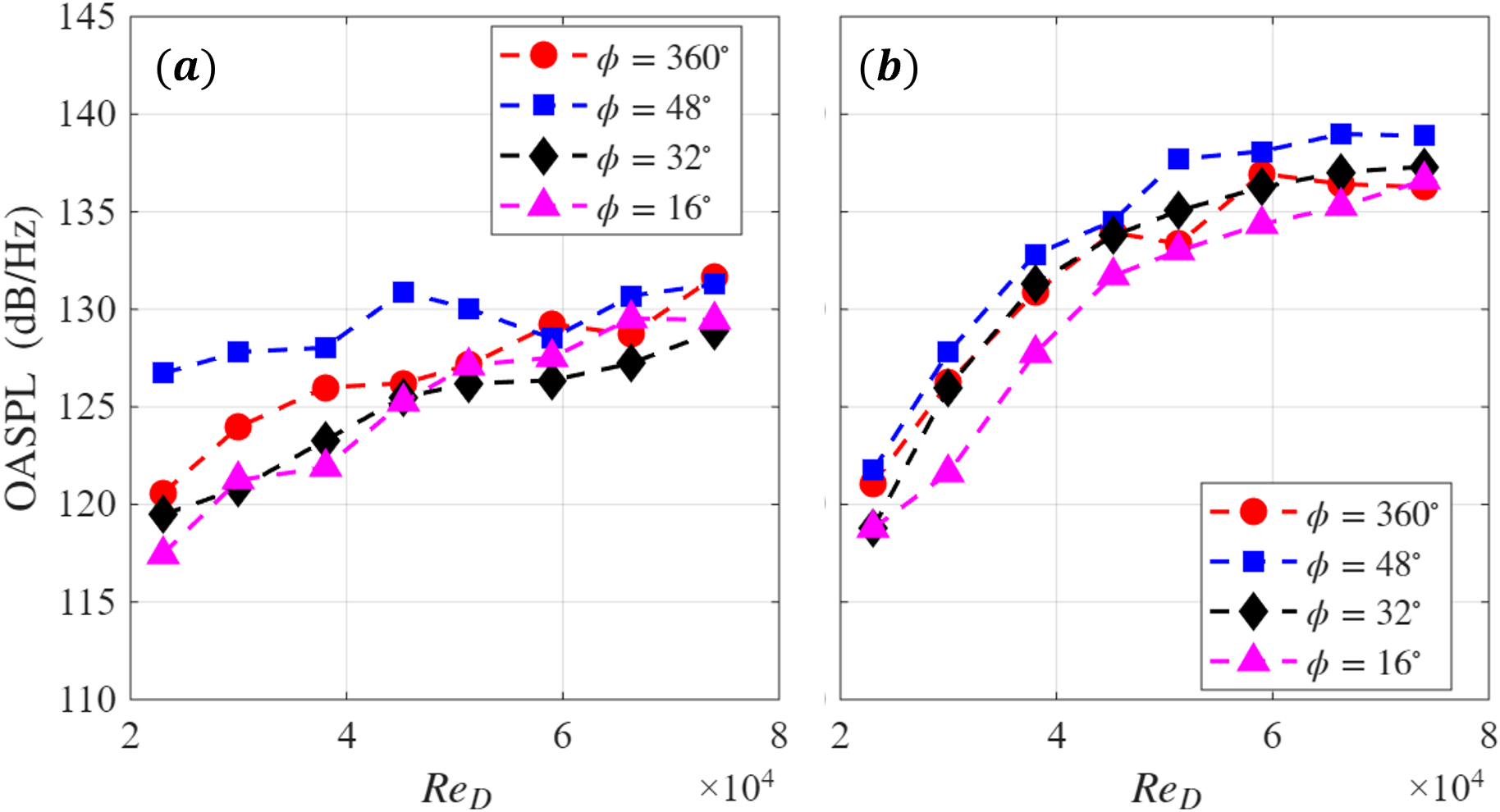}
	\caption{ OASPL variation for different segmentation angles and $Re_D$ on cavity rear wall for (a) $L/D=1.0$  (b) $L/D=6.0$ }
    \label{fig:OASPL_segmentation}
\end{figure*}

Figure \ref{fig:OASPL_segmentation} shows the overall sound pressure level (OASPL) calculated on the cavity floor center $(S_p)_{fc})$for $L/D=1.0$ and $6.0$ at different $\phi$ and $Re_D$s. The OASPL values for $L/D=1.0$ (see Figure \ref{fig:OASPL_segmentation}-a) suggest that, for an axisymmetric cavity with $\phi=360^\circ$, OASPL increases with $Re_D$. For $\phi=48^\circ$, OASPL increases significantly compared to the axisymmetric case at lower $Re_D$; as $Re_D$ increases, OASPL decreases and matches that of the axisymmetric cavity. The further decrease in the $\phi=36^\circ$ and $16^\circ$ OASPL values is slightly smaller than in the axisymmetric case. 

OASPL for $L/D=6.0$ at different $\phi$ angles and various $Re_D$ is shown in Figure \ref{fig:OASPL_segmentation}-b. For an axisymmetric cavity ($ \phi=360^\circ$), OASPL continues to increase as $Re_D$ increases. This increase is attributed to increased shear-layer unsteadiness as $Re_D$ increases. This behavior is consistent with our previous studies \cite {Nanda2026}: as $Re_D$ increases, the K-H vortices in the shear layer strengthen, making the shear layer more unsteady and driving the laminar-to-turbulent transition.  The observed OASPL for $L/D=6.0$ is significantly higher than for $L/D=1.0$. This is because, for $L/D=1.0$, the flow within the cavity is more isolated from the outer flow than at $L/D=6.0$. For the other segmentation angles, the OASPL trend is similar to the axisymmetric case; for $\phi=48^\circ$, the OASPL values increased slightly compared to the axisymmetric case. This may be due to the cavity's 3-D bounding effect. Further decreases in $\phi$ resulted in lower OASPL values than in the axisymmetric configuration. The Lowest OASPL is observed at $\phi=16^\circ$. This decrease may be due to the cavity being more isolated from the outer flow as $\phi$ decreases.  OASPL obtained for $L/D=6.0$ for different $\phi$ angles and various $Re_D$ is shown in Figure \ref{fig:OASPL_segmentation}(b). 

\subsection{Spectral Variation}


\begin{figure*}
	\includegraphics[width=0.8\textwidth]{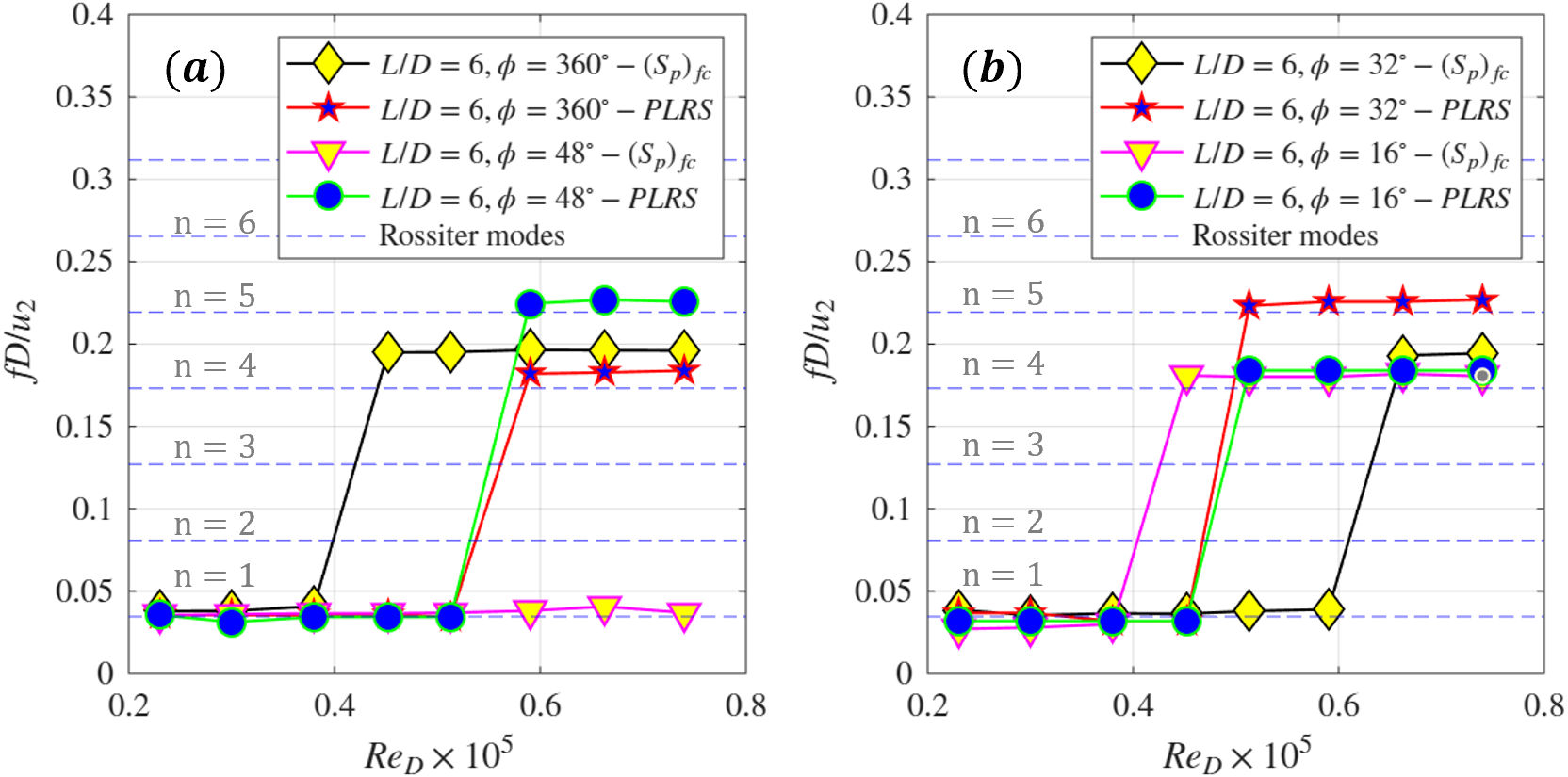}
	\caption{Comparison of the dominant Strouhal numbers $(St_D)$ obtained from the FFT spectra of the normalized fluctuation signals using sensor (
    $(S_p)_{fc})$ measurements and PLRS imaging for $L/D=6$.  (a) $\phi=360^\circ$ and $\phi=48^\circ$ (b) $\phi=32^\circ $and$ \phi=16^\circ$. The dashed blue lines represent the Rossiter modes obtained using the semiempirical relation [Eq. (1)].}
    \label{fig:f_phi_effect_LD6}
\end{figure*}

\begin{figure*}
	\includegraphics[width=0.5\textwidth]{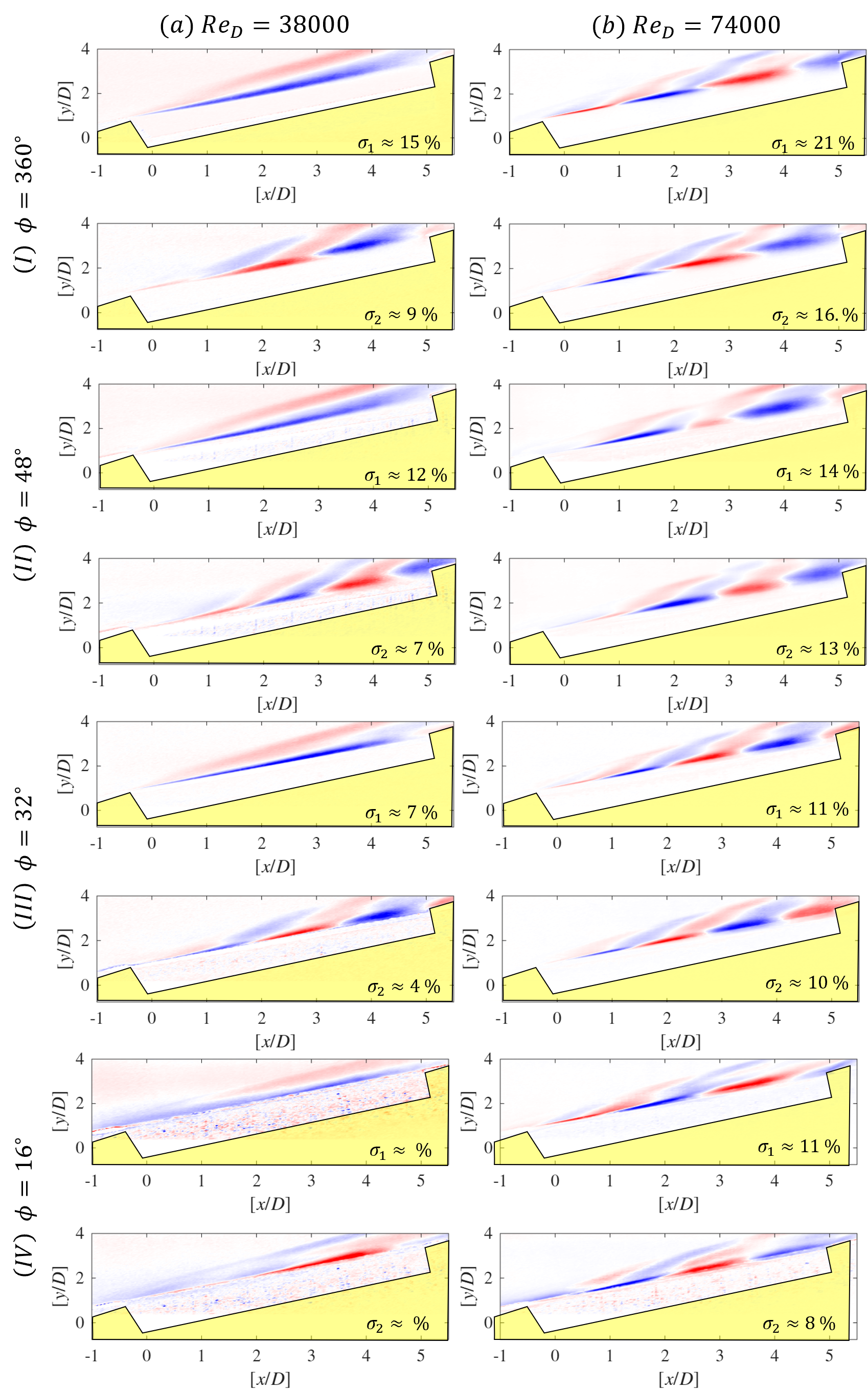}
	\caption{ Proper Orthogonal Decomposition segmented cavities at various $\phi$ angles for (a) $Re_D=38000$ and (b) 74000.}
    \label{fig:POD}
\end{figure*}

\begin{figure*}
	\includegraphics[width=0.35\textwidth]{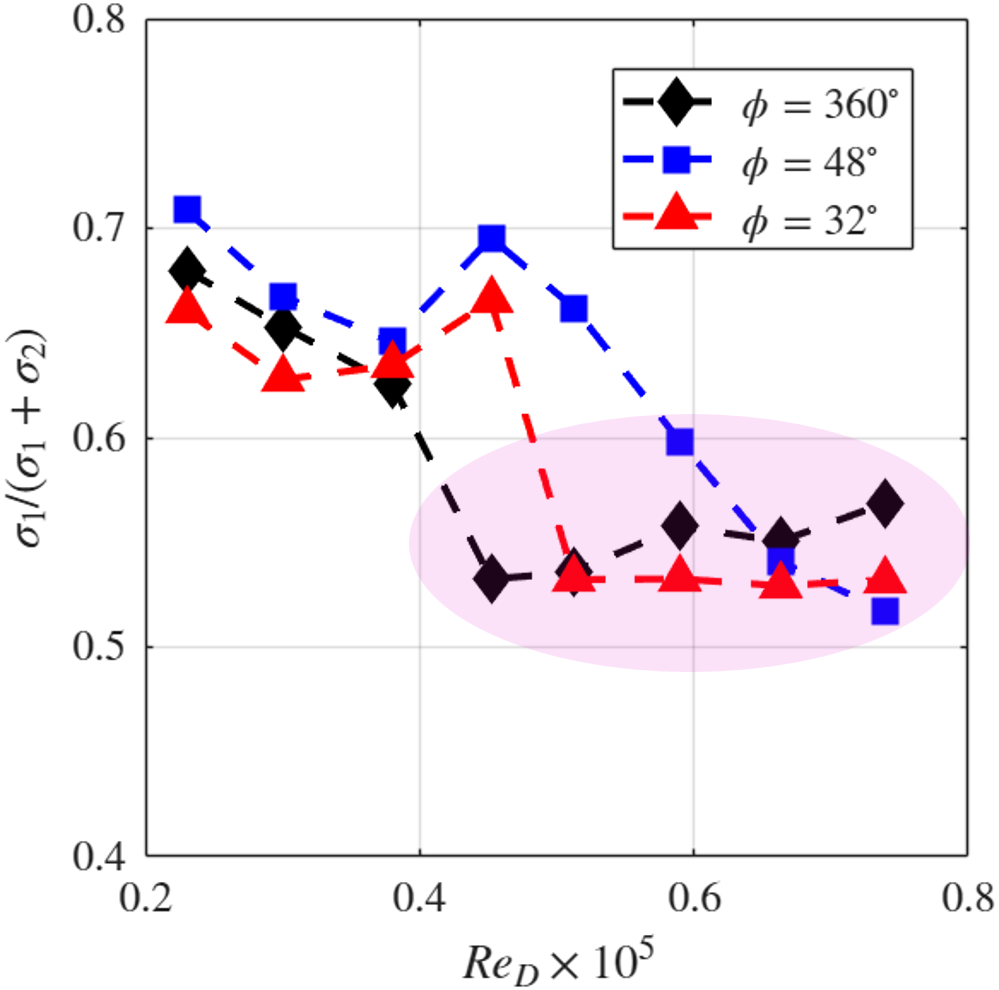}
	\caption{ Comparison of the dominant energy modes ratio for $\phi = 360^\circ$, $48^\circ$, and $32^\circ$ at different $Re_D$ values. $\sigma_1$ and $\sigma_2$ represent the energy fractions associated with the first and second modes, respectively.}
    \label{fig:POD_energy}
\end{figure*}

\subsubsection{Spatial mode identification}

The dominant non-dimensional frequencies for the different cavity segmentation angles are determined from fast Fourier transform (FFT) analysis of the pressure signals, as presented in Figure\ref{fig:f_phi_effect_LD6}. The corresponding PLRS snapshots are also subjected to FFT analysis to identify the dominant frequencies associated with the shear-layer dynamics. The frequencies obtained from the floor pressure measurements and PLRS analysis are compared with the theoretical Rossiter modes.
For the axisymmetric cavity ($\phi=360^\circ$), the dominant non-dimensional frequency ($fD/u_2$) at lower Reynolds numbers is approximately 0.035, corresponding to the first Rossiter mode. Both the floor pressure and PLRS measurements consistently identify this low-frequency component. With increasing $Re_D$, a transition toward higher-frequency oscillations is observed, with the dominant frequency approaching the fourth Rossiter mode ($fD/u_2\approx0.19$). This transition is first captured by the floor pressure measurements at $Re_D\approx45,200$, whereas the PLRS measurements identify it at a slightly higher Reynolds number of $Re_D\approx51,300$. The shift in the dominant frequency indicates a change in the underlying shear-layer dynamics, associated with the modification of the hydrodynamic-acoustic feedback mechanism. At intermediate Reynolds numbers, the low-frequency flapping motion and higher-frequency Kelvin--Helmholtz (K-H) instability may coexist, resulting in competing spectral contributions. Consequently, the dominant frequency identified by the two measurement techniques may differ depending on their respective sensitivity to the flow structures. In particular, the intermediate Rossiter modes (second and third modes) may also remain present in this transitional regime, as discussed by Nanda et al.~\cite{Nanda2026}. Nevertheless, this analysis focuses primarily on identifying the dominant frequency and associated flow instability rather than resolving all spectral components.

Introducing cavity segmentation considerably modifies this transition behavior, with the extent of the modification depending on the segmentation angle. For $\phi=48^\circ$ (see Figure\ref{fig:f_phi_effect_LD6}-a), the floor pressure measurements indicate that the first Rossiter mode remains dominant throughout the investigated Reynolds-number range, suggesting that the cavity feedback response remains governed by the low-frequency flapping motion. In contrast, the PLRS measurements identify a transition toward the fifth Rossiter mode at $Re_D=59,000$. This difference indicates that, although the cavity feedback response remains dominated by the low-frequency component, higher-frequency shear-layer structures become increasingly prominent in the PLRS measurements. A further reduction in the segmentation angle to $\phi=32^\circ$ (see Figure~\ref{fig:f_phi_effect_LD6}-b) results in a more pronounced difference between the two measurement techniques. The PLRS measurements capture an earlier transition toward the fifth Rossiter mode at $Re_D=51,300$, whereas the floor pressure measurements identify a shift toward the fourth Rossiter mode at $Re_D=66,300$. This behavior suggests that the shear-layer dynamics become increasingly sensitive to segmentation, with the convective instability becoming prominent before the corresponding change is reflected in the dominant floor pressure response. The earlier identification of the transition by PLRS may be attributed to its measurement location within the shear layer, which provides a more direct representation of the evolving convective structures, whereas the floor pressure signal reflects the integrated response of the cavity and its acoustic feedback mechanism.When the segmentation angle is further reduced to $\phi=16^\circ$, the transition toward higher-frequency oscillations occurs at a comparatively lower Reynolds number (see Figure\ref{fig:f_phi_effect_LD6}-b). Both measurement techniques identify this transition within a similar Reynolds-number range, with the floor pressure measurements indicating a shift toward the fourth Rossiter mode at $Re_D= 45,200$, followed by the PLRS measurements at $Re_D=51,300$. In this case, both techniques identify the fourth Rossiter mode as the dominant higher-frequency component, indicating improved agreement between the cavity pressure response and the shear-layer dynamics.

\subsubsection{Proper Orthogonal Decomposition (POD)}

To further investigate the spatial characteristics of the dominant flow instabilities identified through the frequency analysis, proper orthogonal decomposition (POD) is applied to the high-repetition-rate PLRS snapshots. The POD methodology follows the approaches described in \cite{lumley1967,sirovich1987,Meyer2007,Taira2017}. A total of 600 snapshots are considered for the decomposition, and the convergence characteristics of the POD analysis are discussed in detail by Nanda et al.\cite{Nanda2026}. Figure\ref{fig:POD} presents the dominant spatial modes and their corresponding energy contributions for the different cavity configurations. At the lower Reynolds number, $Re_D=38,000$, the first dominant POD mode exhibits a flapping-type motion of the separated shear layer for all investigated configurations, including $\phi=16^\circ$, $32^\circ$, $48^\circ$, and the axisymmetric cavity ($\phi=360^\circ$). The alternating red and blue regions represent the compression and rarefaction components of the pressure fluctuations, respectively, and illustrate the large-scale oscillatory displacement of the shear layer. The corresponding second dominant mode exhibits K-H type shedding structures for $\phi=360^\circ$, $48^\circ$, and $32^\circ$, whereas for $\phi=16^\circ$, it retains a flapping-type spatial distribution. These observations indicate that, at lower Reynolds numbers, the large-scale shear-layer flapping remains the principal flow feature, although smaller-scale convective instabilities may also contribute to the flow dynamics. As shown in Figure\ref{fig:POD_energy}, the energy ratio ($\sigma_1/(\sigma_1+\sigma_2)$) associated with the first dominant mode is approximately 65\% of the combined first and second modes for $Re_D\leq38,000$, indicating a relatively strong contribution from the leading mode. At the higher Reynolds number, $Re_D=74,000$ (see Figure\ref{fig:POD}-II), a qualitative change in the dominant spatial structures is observed. For all cavity configurations, both the first and second dominant modes exhibit K-H type shedding structures, characterized by alternating regions associated with the convection of instability waves along the shear layer. Unlike the lower Reynolds-number cases, where the dominant mode is primarily associated with large-scale flapping, the higher Reynolds-number flow is characterized by the development of convective shear-layer instabilities. Furthermore, the energy ratio associated with the first mode decreases to approximately 50\%, indicating a more balanced contribution from the two leading modes. This redistribution of modal energy suggests that the flow becomes increasingly dominated by multiple energetic structures rather than a single large-scale oscillatory mode.

\section{Conclusions} \label{sec:conc}

Experiments are conducted on a hypersonic cone-cavity model in the Hypersonic Ludwieg Tunnel (HLT) at Technion, at a freestream Mach number of $M_\infty=6.0$. The effects of cavity geometry, Reynolds number, and three-dimensional confinement on the mean and unsteady flow characteristics are investigated. The Reynolds number based on cavity depth ($Re_D$) is varied from $23,000$ to $74,000$. To examine the effects of three-dimensional confinement, cavities with segmentation angles of $\phi=16^\circ$, $32^\circ$, and $48^\circ$ are considered. Rayleigh scattering is used to characterize flow-field unsteadiness in the segmented cavities, while unsteady pressure sensors and coaxial surface junction thermocouples (CSJTs), positioned upstream, inside, and downstream of the cavity, quantify the pressure and heat-flux distributions. The major findings of the present study are summarized as follows.

\begin{itemize}

\item For $L/D=1$, irrespective of the segmentation angle, the downstream Strouhal number ($St$) and mean pressure distributions of the segmented cavities closely follow those of the axisymmetric configuration. However, the mean pressure values within the cavity are higher, and the $St$ value is lower for the segmented configurations, indicating the influence of three-dimensional confinement. For $L/D=6$, similar agreement is observed in the mean pressure distributions, and the cavity rear-wall and downstream Strouhal number increase with the segmentation angle. 

\item The FFT analysis reveals that the segmentation angle significantly influences the onset of higher-frequency shear-layer instabilities for $L/D=6$. Reducing the segmentation angle from $\phi=48^\circ$ to $16^\circ$ promotes an earlier transition from low-frequency flapping to higher-frequency shedding. Comparing floor pressure and PLRS measurements further reveals that the identified transition depends on the measurement technique and segmentation angle. While both techniques capture comparable transitions for $\phi=16^\circ$, noticeable differences are observed for $\phi=32^\circ$ and $48^\circ$. This behavior highlights the complementary nature of the two measurement techniques, with PLRS providing spatially resolved information about shear-layer dynamics, while floor pressure measurements characterize the pressure fluctuations at the cavity floor.

\item The POD analysis reveals a Reynolds number-dependent evolution of the dominant spatial flow structures. At lower Reynolds numbers, the flow is primarily characterized by large-scale shear-layer flapping, whereas at higher Reynolds numbers, K-H type shedding structures become dominant. This transition is accompanied by a more balanced distribution of energy between the first two POD modes, indicating an increased contribution from multiple energetic flow structures. These observations are consistent with the FFT analysis, which identifies a shift from the first Rossiter mode toward higher-order Rossiter modes as $Re_D$ increases. The agreement between the frequency spectra and POD spatial distributions provides complementary evidence of the progressive development of shear-layer instabilities and their associated convective structures.

\end{itemize}

\section*{Acknowledgments} 
The authors would like to acknowledge the help of lab members Michael Dunaevsky, Oleg Kan, Nadav Shefer, David Naftali, and Efim Shulman in conducting experiments in the Hypersonic Ludwig Tunnel (HLT) at the Technion Wind Tunnel Complex.


\section*{Declaration of interests}
The authors report no conflict of interest.

\section*{References}
\vspace{0.1cm}
\nocite{*}
\bibliography{references}

\end{document}